\documentclass[sigconf]{acmart}
\usepackage{enumitem}
\usepackage{xspace}
\usepackage[most]{tcolorbox}
\usepackage{booktabs}           
\usepackage{tabularx}           
\usepackage[table]{xcolor}      
\usepackage{hyperref}           
\usepackage{threeparttable}
\usepackage{adjustbox}
\usepackage{makecell}

\AtBeginDocument{%
  }

\copyrightyear{2026}
\acmYear{2026}
\setcopyright{cc}
\setcctype{by}
\acmConference[MSR '26]{23rd International Conference on Mining Software Repositories}{April 13--14, 2026}{Rio de Janeiro, Brazil}
\acmBooktitle{23rd International Conference on Mining Software Repositories (MSR '26), April 13--14, 2026, Rio de Janeiro, Brazil}
\acmPrice{}
\acmDOI{10.1145/3793302.3793344}
\acmISBN{979-8-4007-2474-9/2026/04}

\begin{document}

\title{ML in a Box: Analyzing Containerization Practices in Open Source ML Projects}


\author{Faten Jebari}
\affiliation{%
  \institution{Grand Valley State University}
  \state{Michigan}
  \country{USA}
}
\email{jebarif@mail.gvsu.edu}
\orcid{0009-0002-6205-7590}

\author{Emna Ksontini}
\affiliation{%
  \institution{University of North Carolina Wilmington}
  \state{North Carolina}
  \country{USA}
}
\email{ksontinie@uncw.edu}

\author{Amine Barrak}
\affiliation{%
  \institution{Oakland University}
  \state{Michigan}
  \country{USA}}
\email{aminebarrak@oakland.edu}

\author{Wael Kessentini}
\affiliation{%
  \institution{DePaul University}
  \state{Illinois}
  \country{USA}}
  \email{wkessent@depaul.edu}

\begin{abstract}
Containerization has become increasingly essential in the machine learning (ML) domain, providing reproducibility, portability, and environment consistency. While prior studies have analyzed Dockerfile structures and best practices, none have examined ML projects in depth to reveal how the iterative nature of ML workflows influences container footprint, build performance, and caching behavior.

We present the first large scale empirical study of 1,993 ML related Dockerfiles, combining quantitative analysis of container roles in ML projects and build dynamics with a qualitative investigation of refactoring practices. Results show that containers serve distinct roles across training, inference, and infrastructure. 
Containers are typically large, averaging 10.27~GB in size, and require long build times of about 8.84~minutes. We find that 44.4\% of commits trigger rebuilds, primarily due to context file changes (96.4\%), with experimentation being the main motive behind those commits that initiate rebuilds. Despite partial cache reuse, 71\% of rebuild work is wasted on redundant computation. From stable projects, we identify 7~recurring ML-specific Dockerfile refactoring patterns that improve build efficiency and reduce container footprint.
\end{abstract}

\ccsdesc[500]{Software and its engineering~Software configuration management and version control systems}
\ccsdesc[300]{Software and its engineering~Software maintenance tools}
\ccsdesc[300]{Computing methodologies~Machine learning}
\ccsdesc[200]{Information systems~Empirical software engineering}

\keywords{Containerization, Dockerfile, Machine Learning (ML), Build Performance, Cache, Continuous Integration (CI/CD).}

\maketitle

\section{Introduction}
\label{sec:introduction}

Containerization has become the standard foundation for building, testing, and deploying software systems. Teams use containers to encapsulate dependencies, standardize runtime environments across different operating systems and infrastructures, and simplify CI/CD processes \cite{henkel2020learning}. Compared to heavy virtualization, containers provide lower overhead and faster startup times, making them appealing not only for production microservices but also for developer tools and data pipelines. In practice, container images now serve as the main unit of delivery and the center of automation for much of modern software engineering \cite{Strudel2018FSE, cito2016using}.

Machine learning (ML) projects intensify the same needs, portability, repeatability, and automation, while adding hardware and dependency demands that stress container workflows. Training pipelines must coordinate large framework stacks, accelerator drivers (e.g., CUDA/cuDNN), and toolchains \cite{gonzalez2023container} and serving pipelines must package models for diverse targets (CPU/GPU) \cite{barrak2025cost, riahi2020comparison}. Prior empirical work on ML projects reports the broad and routine use of Docker across various types of ML applications, such as MLOps, AIOps, toolkits and deep learning frameworks and identifies many distinct purposes for containerization, including cross-OS portability, runtime alignment, and CPU/GPU selection. \cite{Openja2022EASE,houerbi2024empirical}. At the same time, these studies caution that ML images tend to be resource-hungry due to numerous layers and deeply nested file trees, reflecting the heavy artifacts (framework binaries, models, data) typical of ML practice \cite{LinICSME2020,Cito2017MSR,ksontini2021refactorings}.

Moreover, ML development is inherently iterative and experimental, involving continual cycles of model selection, hyperparameter tuning, and feature engineering, each requiring frequent code and configuration changes to achieve optimal performance \cite{amershi2019software, xin2018developers}. This rapid iteration cycle directly interacts with container build mechanics \cite{houerbi2024empirical}. Docker builds, however, are linear and layered, image are constructed by executing the Dockerfile top-to-bottom, materializing one layer per instruction. Cache lookup proceeds stepwise: for each instruction, the builder searches for a previously produced layer whose cache key matches the current state (parent image/layer ID, the instruction text and relevant arguments, and for \texttt{COPY/ADD}, the content digests of referenced files in the build context). If the key matches, the cached layer is reused; otherwise, the cache “breaks,” and the current instruction and all subsequent instructions are re-executed, even if they themselves are unchanged. Under these semantics, rebuilds are triggered by two classes of changes: edits to the Dockerfile itself and edits to any file copied inside the container \cite{dockercaching, rosa2025mining}. This works well for stable, late-changing build recipes, but it is fragile under ML’s edit patterns, where minor changes can trigger costly rebuilds of heavy installation steps.

Despite widespread container use in ML and mounting evidence of large, complex images, we lack a systematic, role-aware account of four fundamentals: (i) what functional roles containers actually serve within ML projects; (ii) how large ML containers are in the wild and how long cold builds take; (iii) how frequently rebuilds occur over project histories, what triggers them (Dockerfile edits versus context changes), and how much prior work the cache truly preserves once invalidated; and (iv) which behavior-preserving Dockerfile refactorings practitioners employ and how those patterns align with observed footprint and rebuild pain points. Prior work at the ML–Docker intersection describes where containers are used \cite{Openja2022EASE} and repository-level studies catalog general Docker practices and fragility \cite{Cito2017MSR, Ksontini2025MSR}, while registry-scale measurements highlight redundancy and limited natural layer sharing \cite{Zhao2019Cluster}. What is missing is a quantitative, role-sensitive picture that links image footprint, rebuild mechanics, and concrete refactoring behavior in ML projects.

This paper addresses that empirical gap. We analyze how ML Dockerfiles evolve, how code and dependency edits interact with caches, and which layout choices reduce rebuild pain, linking developer intent to rebuild cost and cache reuse. We report the following contributions:

\begin{itemize}

    \item We present a container role taxonomy and large-scale measurement of ML containers, classifying 1993 Dockerfiles out of 392 different ML projects and quantifying footprint and cold-build time per role, including base versus custom layer contributions and failed-build overhead.
    \item We provide an emperical analysis of rebuild dynamics over 1.06M commits that measures how often rebuilds occur, what triggers them, and how much prior work is actually reused once the cache breaks.
    \item We show how developer intent (experiments, dependency/env, infrastructure/CI, bug fixes) shapes cache outcomes, explaining role-specific fragility.
    \item We curate a catalog of 7 Ml-specific Dockerfile refactorings grounded in real commits.
\end{itemize}

\textbf{Replication Package.} All material, prompts and data used in our study are available in our replication package~\cite{replication}.

\section{Related Work}
\label{sec:related_work}
Our research bridges foundational Docker ecosystem studies and containerized ML workflows, highlighting gaps in applying Docker practices to ML development.

\subsection{Docker Ecosystem Foundations}
Early research into Docker, a cornerstone of modern software deployment, sought to characterize its use in open-source projects. A foundational study by Cito \textit{et al.} \cite{Cito2017MSR} mined 70,000 GitHub Dockerfiles, identifying common practices and quality issues like widespread "code smells," which are violations of best practices that cause bloated images and faulty builds. Complementing this, Ksontini \textit{et al.} \cite{ksontini2021refactorings} examined refactorings and technical debt in Docker projects, revealing that developers often perform corrective and preventive refactorings to address maintainability and build issues.

Subsequent research focused on these “Dockerfile smells,” or suboptimal practices impairing performance and security. Henkel \textit{et al.} developed the \textit{Binnacle} tool to automatically detect and fix smells by mining improvement rules from Dockerfile histories, noting that community files violated best practices more than expert-written ones \cite{Henkel2020ICSE}. Similarly, Lin \textit{et al.} analyzed the ecosystem's evolution, observing a gradual reduction in smells and a shift toward smaller base images over time \cite{LinICSME2020}.

Building on this, researchers have worked to automate smell correction. Rosa \textit{et al.} demonstrated that automatically fixing pervasive smells, such as improper package pinning, significantly reduces image bloat and build times \cite{Rosa2024EMSE}. Their evaluations also revealed that not all smells carry equal weight; some are minor, while others are strongly correlated with negative outcomes \cite{Rosa2024MSR}.

The research scope has also expanded beyond Dockerfiles to the broader ecosystem. Xu and Marinov advocated mining container image repositories like Docker Hub, arguing that images offer richer runtime insights than source code alone \cite{Xu2018ICSE}. Supporting this view, Schermann \textit{et al.} constructed a dataset of Dockerfile evolution to enable longitudinal analysis of container configurations \cite{Schermann2018Docker}.

Most recently, machine learning has been applied to automate Dockerfile improvement. Addressing the tendency for developers to postpone refactoring, which increases image size and build times, Ksontini \textit{et al.} used large language models for automated refactoring \cite{Ksontini2025MSR}. Their In-Context Learning approach generated improved Dockerfiles without fine-tuning and significantly outperformed manual efforts in reducing image size and build duration.

\subsection{Containerizing Machine Learning Workflows}
Moving models from experimentation to operation raises recurring concerns about reproducibility, auditability, and release cadence. Reports describe environment drift across laptops, build servers, and clusters, which complicates handoffs between data science and platform engineering. Containers address these concerns by providing portable and versioned environments, controlled rollback, and a clear interface between roles. Owen \textit{et al.} distill these needs into concrete guidance on Dockerfiles, image versioning, and CI/CD integration, presenting containerization as the connective tissue between development and operations for machine learning \cite{Owen2025Containerization, barrak2025faasguard}.

This principle has been widely adopted in machine learning, prompting recent software engineering research to investigate how ML projects specifically leverage containerization. Houerbi’s empirical study \cite{houerbi2024empirical} of continuous integration and delivery in open-source ML projects shows that pipelines frequently use containers to standardize job environments, reinforcing containerization as a key enabler of repeatable automation in ML workflows. Recent work on the co-evolution of ML pipelines and source code further shows that changes to models, data, and configurations often accompany code edits, highlighting the iterative and experiment-driven nature of ML development and the importance of reproducible environments \cite{barrak2021co}.
Complementing this pipeline level perspective, Openja \textit{et al.} \cite{Openja2022EASE} examined 406 open-source ML-based projects on GitHub that also host corresponding Docker images on Docker Hub. They found that container usage is pervasive across the ML landscape, from end-user applications and MLOps tools to core frameworks like TensorFlow and PyTorch. Their analysis shows that ML projects often require highly specific and difficult-to-configure runtime environments, including exact versions of CUDA drivers and Python packages. However, their analysis considered all Dockerfiles within ML repositories without distinguishing whether they served ML workflows or auxiliary purposes, such as web frontends or databases.

This paper presents the first large-scale empirical study linking ML development activities to Docker build times, image footprint, rebuild frequency, and cache fragility, revealing current practices and ML-specific refactoring patterns.
\begin{figure*}[t]
\centering
\includegraphics[width=0.98\linewidth]{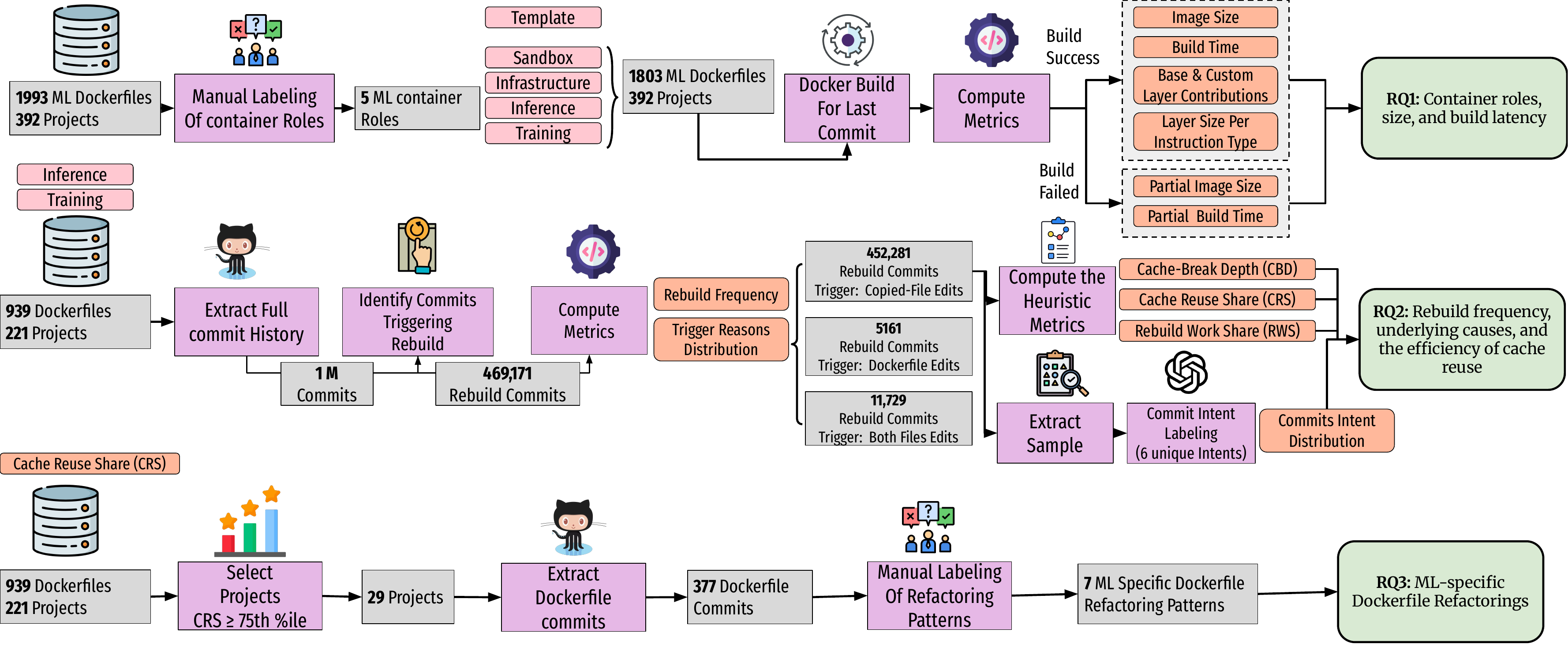}
\caption{Overview of the Study Design}
\label{fig:approach}
\end{figure*}

\section{Study Design}
\label{sec:study-design}
Figure~\ref{fig:approach} presents an overview of our study design, outlining the multi-stage analysis conducted to answer the three research questions. The workflow begins with identifying the functional roles of containers in ML repositories and analyzing their image structure, size, and build time (\textbf{RQ1}). It then quantifies rebuild frequency and cache reuse efficiency across projects (\textbf{RQ2}) and concludes with the identification of ML-specific Dockerfile refactoring patterns to optimize build performance and image footprint (\textbf{RQ3}).

\subsection{Research Questions}

\textbf{RQ1. What roles do containers serve in ML repositories, and how do these roles differ in image footprint and build-time latency?}

\underline{\textit{Motivation.}} Containerization is now common in the ML lifecycle, with repositories often managing several containers \cite{Openja2022EASE}. Previous work has demonstrated Docker’s importance in ML ecosystems and described the types of systems that utilize it \cite{Openja2022EASE}. However, we still lack an empirical account of how ML projects differentiate and organize their containers. The structural and operational diversity of containers that support ML development remains undercharacterized. ML containers routinely bundle deep learning frameworks, hardware acceleration runtimes, and sizable model or data artifacts. Such stacks inflate image size and complicate builds, and practitioners routinely report slow feedback cycles in CI/CD and local experimentation \cite{houerbi2024empirical}. Yet, despite these reports, there is no data-driven baseline that relates concrete types of ML containers to their build-time behavior and image footprint. This research question aims to identify the role containers play in ML repositories and to empirically measure their image size, build latency, and structural composition, thereby providing a descriptive baseline of current practices.

\underline{\textit{Approach.}} To address RQ1, we analyze 1,993 ML-related Dockerfiles as container specifications and classify each into a container role based on file-level evidence, augmented with repository-level context. 
Four authors conduct labeling in three steps: first, an 80-item stratified calibration batch (all four authors label all 80) aligns decision rules for inferring type from observable cues (e.g., \texttt{FROM} base image and stage names, \texttt{ENTRYPOINT}/\texttt{CMD} targets, invoked scripts, dependency stacks, inline comments, and build context); second, a main pass over the remaining 1,913 files in which 25\% (478 files) are double-coded by rotating pairs and the remainder (1,435) are single-coded, yielding 956 labels on the double-coded pool; and third, we meet to resolve disagreements and agree on a final label for each file. For the double-coded Dockerfiles, the pairwise Cohen’s K ranged from 0.72 to 0.83, indicating a solid level of pairwise reliability. On the calibration set, Fleiss’ K = 0.78.

For measurement, we build exactly the last committed version of each Dockerfile in our dataset. Successful builds provide the resulting image size and wall-clock build time; we perform three independent rebuilds and report the average duration across the three runs. When a build fails, we retain partial evidence by recording the elapsed time and the total size of layers produced up to the point of failure. For consistency, we repeat the attempt three times and calculate the average of the partial durations. Additionally, for each image, we calculate the proportion of the total footprint that the base image accounts for. To describe structural characteristics, we record the types of custom layers and, among the non-base layers, find the typical (median) layer size within each image. Finally, we group all these measurements by container role. We note that all experiments were conducted on an Ubuntu-based workstation equipped with an AMD Ryzen 7 3700X 8-core CPU, 32 GB RAM, and an NVIDIA GeForce RTX 2060 SUPER GPU.

\textbf{RQ2. How frequently do container rebuilds occur in ML projects, to what extent is cached work reused, and which types of commits typically trigger these rebuilds?}

\underline{\textit{Motivation.}} RQ1 established a baseline: ML containers are large artifacts, and their builds take a significant amount of time. However, these figures only tell part of the story. ML development is usually iterative, involving quick cycles of refining code, configurations, and experimental setups \cite{amershi2019software}. The real practical cost for teams depends less on the initial expense of creating a container and more on how often this cost occurs during a project and what types of changes trigger it. Docker's layer cache can prevent recomputing work, but reuse depends on what changed and where in the Dockerfile the change happens; when a layer's inputs change, that layer and all subsequent layers are rebuilt. What we lack is a commit-level view: how often images are rebuilt, how much work is reused, and which commit intents tend to invalidate the cache.

This research question aims to quantify rebuild frequency and cache reuse at commit granularity, and to identify the intents that drive invalidation.

\underline{\textit{Approach.}}
For RQ2, we limit our analysis to \emph{training} and \emph{inference} containers (as labeled in RQ1). These artifacts are part of the ML critical path, contain heavyweight stacks (such as frameworks and accelerator runtimes), and show well-defined change types (e.g., updates to training pipelines, serving code, or models). Therefore, we exclude: (i) \emph{infrastructure} Dockerfiles used to provision or configure systems, since their rebuild timing depends on topology; (ii) \emph{sandbox} images that lack a specific entrypoint or have unclear intent; and (iii) \emph{templates}, which are not meant to be built as-is.

We analyze 939 Dockerfiles (632 inference, 307 training) from 221 repositories. For each Dockerfile, we examine the repository history starting from the initial commit that introduces it and identify a rebuild-triggering commit whenever either (i) the Dockerfile changes or (ii) a file copied into the image via \texttt{COPY} or \texttt{ADD} or \texttt{RUN --mount=type=bind} is modified; for (ii), we determine eligible files after applying the project’s \texttt{.dockerignore} rules to the build context. Using this event stream, we first report the rebuild frequency based on container role and then compare which change type triggers rebuilds more often (Dockerfile edits versus copied-file edits). 

To study caching, we focus solely on commits where the copied files changed but the Dockerfile itself did not. In these cases, we estimate cache reuse (and the associated rebuild work) by comparing each triggering commit to its immediate predecessor, assuming a hot cache (i.e, the best-case scenario). Since rebuilding every historical commit is impractical at our scale (total of 452{,}281 copied-file edits commits ) and could introduce noise from machine, network, and registry factors, we employ a static heuristic aligned with Docker’s cache semantics (\autoref{sec:heuris}) to infer how much work would be reused versus re-executed under optimal caching conditions. We note that the introducing commit is treated as a full rebuild with no previous cache. We linearize history on the default branch (first-parent) to prevent double-counting merges and ignore changes to files not referenced by the Dockerfile, as they cannot influence its build.

Moreover, we classify commits into six intent categories adapted from prior work and refined for ML development contexts \cite{houerbi2024empirical,zampetti2021ci}:

\begin{itemize}[leftmargin=1.2em]
    \item \textbf{Experiment:} changes to models, hyperparameters, or datasets.
    \item \textbf{Feature or Enhancement:} new features, refactoring, cleanup, performance, or security improvements.
    \item \textbf{Bug Fix:} corrective maintenance and error correction.
    \item \textbf{Dependency / Environment:} updates to packages, frameworks, CUDA drivers, base images, or environment configuration.
    \item \textbf{Infrastructure or CI:} build, testing, or deployment tooling.
    \item \textbf{Documentation:} nonfunctional or descriptive edits.
\end{itemize}

We draw a stratified random sample by \emph{container role} (RQ1) and by \emph{repository}. For each role, let \(N_r\) denote the number of cache invalidating commits in repository \(r\); we allocate a target number to each repository using weights \(w_r=\sqrt{N_r}\). We then sample the assigned number uniformly without replacement within each repository using a fixed random seed. This design preserves proportional coverage while preventing large repositories from dominating. The final sample contains 2990 commits, which we label with a LLM (OpenAI GPT\mbox{-}4o). The model receives the commit message and the file-level diffs that can invalidate Docker’s cache in that commit. The prompt requires exactly one primary intent from the six categories and, when the evidence supports it, allows one secondary intent. We also allow an \emph{Uncategorized} option when the available evidence is insufficient or contradictory, which avoids speculative assignments and reduces label noise. To assess label quality, we uniformly sample \textbf{50} commits per intent category (\textbf{300} total) and have an author independently relabel them using the same evidence. We report primary exact match, the share of case where the author’s primary equals the model’s primary (79\%), and set overlap, the share where the author’s label set (one or two intents) intersects the model’s label set (89\%).

\textbf{RQ3. Which ML-specific Dockerfile refactorings are used in practice to accelerate container builds and reduce footprint?}

\underline{\textit{Motivation.}}
RQ1 shows that ML containers are large and slow to build, and RQ2 shows that rebuilds recur but vary widely in how much work the cache reuses. The remaining task is to understand the specific ML-Dockerfile practices that lead to fast builds and smaller footprint in practice. 

This research question investigates the ML-specific Dockerfile refactorings that enable faster rebuilds and smaller images.

\underline{\textit{Approach.}} We hypothesize that repositories with consistently high cache reuse, indicated by a high median Cache Reuse Share (CRS) at the repository level, are more likely to have adopted effective practices. Therefore, we select 29 projects whose median CRS is at or above the seventy-fifth percentile and examine all 377 commits that modify Dockerfiles to identify recurring ML-specific refactorings. We assess their effects on both rebuild efficiency and image size, recognizing that reducing the number of bytes produced can shorten cold build times and decrease work after a cache break. Three authors independently review each commit (including the commit message and Dockerfile diff) to determine whether it qualifies as an ML-specific Dockerfile refactoring. We evaluated agreement at two levels before reconciling the labels. First, for refactoring detection (yes or no for each commit), we calculated Fleiss’ kappa across the three raters, resulting in K = 0.77, with pairwise Cohen’s kappas ranging from 0.72 to 0.83, indicating substantial agreement. Second, for assigning refactoring pattern families, done only on commits where at least two raters agreed that a refactoring exists, Fleiss’ kappa was 0.68. We then conducted consensus meetings: detection differences were first resolved to a single yes or no label; for commits labeled yes, pattern family and effect tags were unified into a single set of gold labels for each commit. Finally, we clustered the confirmed labels into recurring patterns, retaining only those observed in at least two repositories.

\subsection{Data Collection}

We build on the corpus introduced by Idowu et al.\ \cite{idowu2024large}, using their catalog of 31{,}066 Python-based ML repositories as our starting point. We first verified that each repository was public and accessible, yielding 4{,}063 candidates. To capture contemporary practice, we required recent maintenance activity (last updated in early 2024), which retained 2{,}419 projects. We then restricted attention to repositories annotated by the source dataset as covering an \emph{end-to-end} ML pipeline, specifically, all six stages: \emph{Acquisition}, \emph{Preparation}, \emph{Modeling}, \emph{Training}, \emph{Evaluation}, and \emph{Prediction}. We adopt this criterion to focus on production-oriented projects where containerization is used beyond ad-hoc experimentation and to ensure measurements are comparable across repositories rather than dominated by single-stage examples. Next, we excluded educational repositories (tutorials, courseware), identified via an author-curated keyword screen over names and descriptions (full list in \cite{replication}), leaving 2{,}300 projects. Finally, we required at least one Dockerfile per project, yielding the study set of 625 repositories.

From these repositories we collected 2{,}543 Dockerfiles and discarded syntactically invalid files, leaving 2{,}517 Dockerfiles. To separate ML-related containers from clearly non-ML ones at scale, we use a state-of-the-art model (OpenAI GPT-4o), with a fixed prompt over the Dockerfile text and immediate path context; this produced two bins: 2{,}074 ML-related and 443 non-ML. Because all ML-related files are manually typed in RQ1, we validated only the exclusion: we drew a random sample of 200 Dockerfiles from the non-ML bin, had two authors independently judge ML vs.\ non-ML using only the Dockerfile and path context, and reconciled differences. In this audit, 198/200 (99\%) were indeed non-ML, indicating that the screen’s ML/non-ML ranking is reliable for our setting and that the risk of excluding ML Dockerfiles is small.

\subsection{Heuristic: Commit-Level Rebuild Cost}
\label{sec:heuris}

To estimate commit-level rebuild cost, we apply a static heuristic only to context-only triggers from RQ2, that is, consecutive commits \((t{-}1 \rightarrow t)\) where the Dockerfile is unchanged and at least one source referenced by \texttt{COPY} or \texttt{ADD} or \texttt{RUN --mount=type=bind} has changed after applying the project’s \texttt{.dockerignore}. For each eligible pair, we walk the unchanged Dockerfile’s \(N\) instructions and identify the first invalidated instruction \(b\): the earliest \texttt{COPY}/\texttt{ADD}/\texttt{RUN --mount=type=bind} whose inputs changed. Under a hot-cache assumption, all steps before \(b\) are treated as reused, and \(b\) and all subsequent steps are treated as rebuilt. This yields a single breakpoint per pair without executing builds.

From \(b\) and \(N\), we derive two complementary indicators. The first type captures \emph{position}, how far into the build the cache break occurs, expressed as the \textbf{Cache-Break Depth (CBD)}:
\[
\mathrm{CBD} = \frac{b-1}{N}.
\]
A higher CBD indicates that a greater portion of the Dockerfile was reused before the cache broke.

The second type captures \emph{magnitude}, how much work must be re-executed once the cache breaks. Each instruction \(i\) is assigned a structural weight \(w_i\) reflecting its rebuild cost. We compute the \textbf{Cache Reuse Share (CRS)} as
\[
\mathrm{CRS} = 1 - \frac{\sum_{i=b}^{N} w_i}{\sum_{i=1}^{N} w_i},
\]
where \(\mathrm{CRS}=1\) means full reuse and \(\mathrm{CRS}=0\) means a complete rebuild. Its complement, the \textbf{Rebuild Work Share (RWS)} \(=1-\mathrm{CRS}\), represents the fraction of total work re-executed after the cache break.

Instruction weights are defined to approximate rebuild effort, assigning a higher cost to instructions that typically incur greater recomputation and a lower cost to lightweight configuration steps. Reinstalling or fetching dependencies (package managers, VCS clones) is usually the most dominant operation, combining network transfer, solver work, and large unpacking, so it receives the highest baseline. \texttt{COPY}/\texttt{ADD}/\texttt{RUN --mount=type=bind} comes next: it forces hashing and unpacking of potentially significant inputs and often feeds later steps. Other \texttt{RUN} commands are typically lighter transforms, and pure configuration (\texttt{ENV}, \texttt{WORKDIR}, \texttt{CMD}, \texttt{ENTRYPOINT}) is the cheapest. This ordering aligns with Docker guidance \cite{dockercaching} and with the layer characteristics observed in our RQ1 traces:

\[
b_i =
\begin{cases}
1.0, & \text{if } i \in \mathtt{RUN}_{\text{install/clone}},\\
0.6, & \text{if } i \in \mathtt{COPY}/\mathtt{ADD}/\mathtt{RUN}_{\text{--mount=type=bind}},\\
0.4, & \text{if } i \in \mathtt{RUN}_{\text{other}},\\
0.2, & \text{if } i \text{ is configuration }(\mathtt{ENV},\ \mathtt{WORKDIR},\mathtt{ENTRYPOINT}, etc).
\end{cases}
\]

We exclude \texttt{FROM}, as a base-image change forces a full rebuild (\(\mathrm{CBD}=0\), \(\mathrm{CRS}=0\)). To account for variability in dependency size and copy volume, we scale weights by normalized magnitudes:
\[
w_i =
\begin{cases}
b_i(1+p_i), & i \in \mathtt{RUN}_{\text{install/clone}},\\[3pt]
b_i(1+s_i), & i \in \mathtt{COPY}/\mathtt{ADD}/\mathtt{RUN}_{\text{--mount=type=bind}},\\[3pt]
b_i, & \text{otherwise},
\end{cases}
\]
where \(p_i\) and \(s_i\) represent the normalized dependency count and copied file size, capped at the project-level 95th percentile.

\section{Results}

\begin{table*}[t]
\centering
\caption{Roles of Containers in ML Projects and Their Distribution}
\label{tab:dockerfile_distribution}
\small
\begin{adjustbox}{width=0.75\textwidth}
\begin{tabular}{p{2.5cm} p{8.6cm} r r}
\toprule
\textbf{Container Role} & \textbf{Definition} & \textbf{Count} & \textbf{Percentage} \\
\midrule
Training & A container whose primary purpose is to run training or fine-tuning jobs. & 307 & 15.4\% \\
\midrule
Sandbox & An interactive working environment with no fixed entrypoint. It starts in an idle state and is intended for developers to attach and run ad-hoc commands for training, inference, data exploration, or debugging. & 581 & 29.2\% \\
\midrule
Infrastructure & A container used to provision and manage supporting infrastructure (e.g., Kubernetes clusters, cloud resources, and CI/CD pipelines). & 283 & 14.2\% \\
\midrule
Inference & A container whose primary purpose is to serve a trained model as a long-running process. & 632 & 31.7\% \\
\midrule
Template & A reference Dockerfile meant for guidance or copy-and-adapt use, containing placeholders or illustrative steps. & 190 & 9.5\% \\
\midrule
\textbf{Total} & & \textbf{1,993} & \textbf{100.0\%} \\
\bottomrule
\end{tabular}
\end{adjustbox}
\end{table*}

\subsection{RQ1: Container roles, size, and build latency}
Manual review of 2,074 Dockerfiles identified 81 LLM false negatives (non-ML Dockerfiles incorrectly classified as ML). The remaining 1,993 ML-related Dockerfiles, originating from 392 repositories, are categorized into five mutually exclusive functional roles, listed in \autoref{tab:dockerfile_distribution}. 

\textbf{Container use in ML repositories falls into two groups: task-oriented containers (Training and Inference) for model workloads, and support containers (Sandbox and Infrastructure) for development and operations.} Inference (31.7\%) and Training (15.4\%) together account for about half of all Dockerfiles, indicating that many images are built to run models or to produce them. The higher share of inference likely reflects common practice, a single trained model is often packaged into multiple inference images for different environments (e.g., CPU, GPU, or runtime stacks). Sandbox images (29.2\%) form nearly a third, suggesting teams keep interactive, “work-at-the-console” environments alongside task-specific images. Infrastructure images (14.2\%) capture containers used for provisioning, tooling, and operational tasks rather than model execution. Finally, templates (~10\%) are not meant to be executed; they serve as scaffolds or starting points to standardize setups, so we exclude templates from subsequent analyses.

\begin{figure}[H]
\centering
\includegraphics[width=0.95\linewidth]{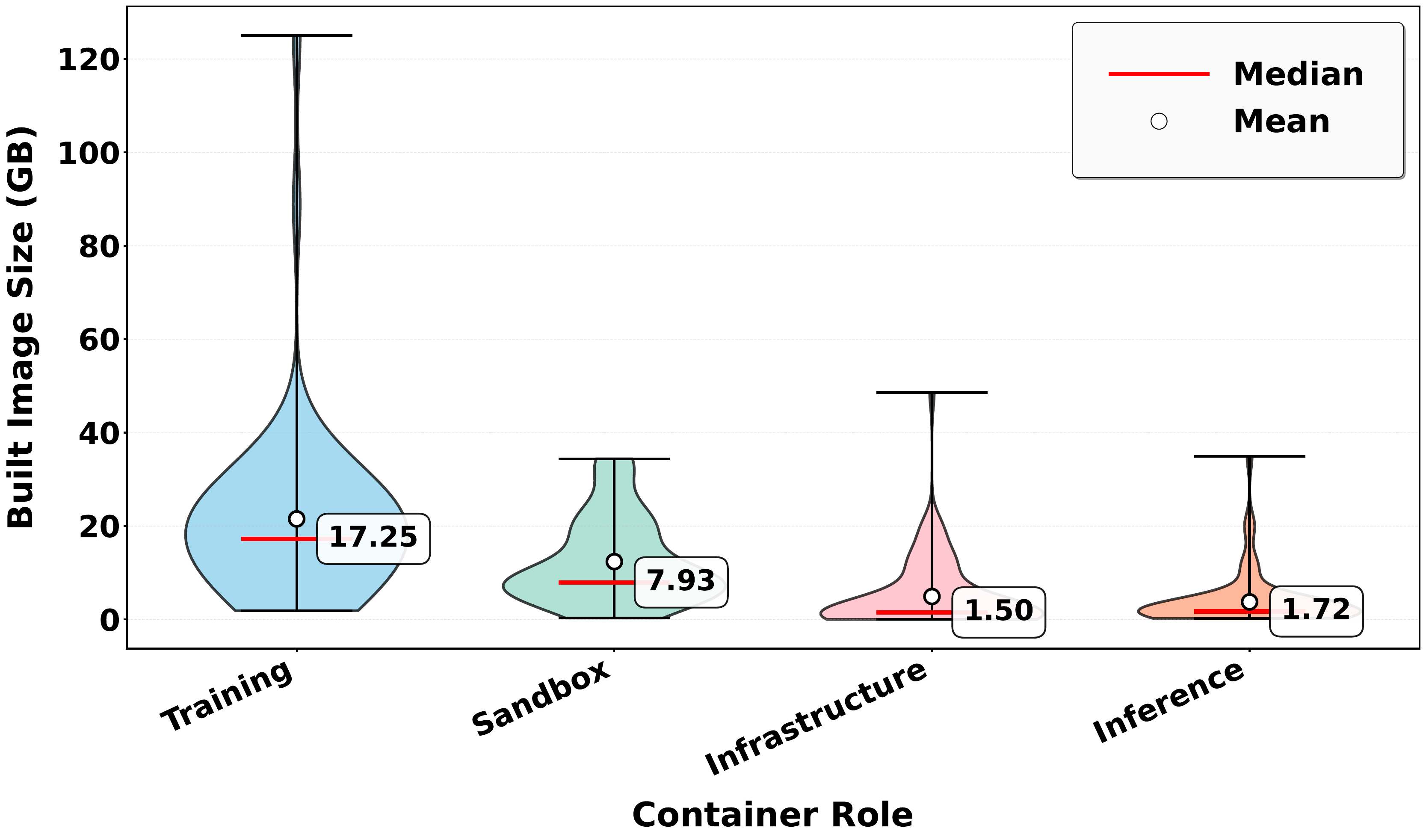}
\caption{Distribution of Container Image Sizes (Fully Built)}
\label{fig:rq1_image_sizes}
\end{figure}

Out of the 1,803 Dockerfiles analyzed (excluding templates), 41\% (739) built successfully while 59\% (1,064) failed. For the successfully built images, we report their fully rebuilt sizes to establish a baseline for footprint across roles (\autoref{fig:rq1_image_sizes}; \autoref{tab:image_size_build_time_combined}).

\textbf{Across container types, images are multi-GB; training is often about 10× larger than Inference or infrastructure, with sandbox in between}. Training containers are the largest, with a median size of 17.25 GB and a long upper tail reaching 125 GB. This reflects the heavy frameworks, GPU drivers, and dataset fragments typically required for training workloads. Sandbox images follow with a median of 7.9 GB, suggesting that interactive environments often package full ML stacks for flexibility. Infrastructure and Inference containers are substantially smaller, with medians of 1.5 GB and 1.7 GB, respectively. These sizes are consistent with lightweight operational tools and serving stacks that only require trained model artifacts and minimal runtime dependencies. The distributions in \autoref{fig:rq1_image_sizes} suggest that the majority of Dockerfiles yield images near or below the median for their role. 

\begin{figure}[H]
\centering
\includegraphics[width=0.95\linewidth]{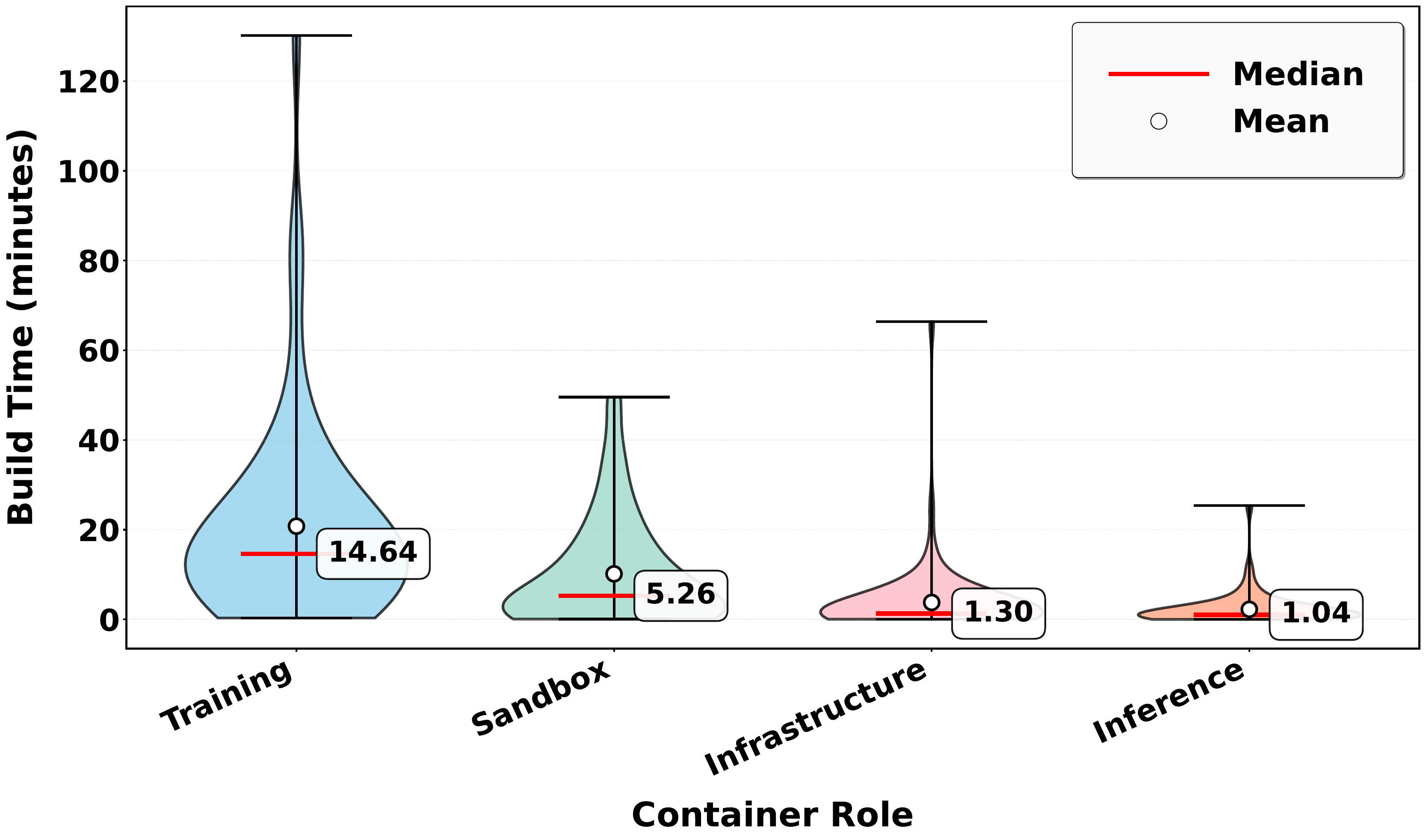}
\caption{Distribution of Image Build Time (Fully Built)}
\label{fig:rq1_image_time}
\end{figure}

\textbf{Builds are generally slow, and build time scales with image size across container types.} Using the same set of successful builds (\autoref{fig:rq1_image_time}; \autoref{tab:image_size_build_time_combined}), Training is slowest (median 14.6 min; max > 2 h), Sandbox is midrange (median 5.3 min), and Infrastructure and Inference are fastest (medians 1.3 and 1.0 min). The ordering mirrors the size results: heavier stacks take markedly longer to build, while inference and infrastructure images complete quickly.

\begin{table}[b]
\centering
\scriptsize
\setlength{\tabcolsep}{3pt}
\caption{Container Image Size \& Build Time (Fully Built)}
\label{tab:image_size_build_time_combined}
\begin{tabular}{lrrrrrrrr}
\toprule
& \multicolumn{4}{c|}{\textbf{Image Size (GB)}} & \multicolumn{4}{c}{\textbf{Build Time (min)}} \\
\cmidrule(lr){2-5} \cmidrule(lr){6-9}
\textbf{Container Role} & \textbf{Median} & \textbf{Mean} & \textbf{Min} & \textbf{Max} & \textbf{Median} & \textbf{Mean} & \textbf{Min} & \textbf{Max} \\
\midrule
Training        & 17.25 & 21.53 & 1.85 & 125.00 & 14.63 & 20.79 & 0.31 & 130.14 \\
Sandbox         & 7.93  & 12.38 & 0.30 & 34.40  & 5.26  & 10.16 & 0.07 & 49.57 \\
Infrastructure  & 1.50  & 4.92  & 0.01 & 48.60  & 1.30  & 3.78  & 0.04 & 66.41 \\
Inference       & 1.72  & 3.76  & 0.23 & 34.90  & 1.04  & 2.25  & 0.00 & 25.37 \\
\midrule
\textbf{Overall} & \textbf{5.69} & \textbf{10.27} & \textbf{0.01} & \textbf{125.00} & \textbf{2.61} & \textbf{8.84} & \textbf{0.00} & \textbf{130.14} \\
\bottomrule
\end{tabular}
\end{table}

\textbf{Failed builds create a slow feedback loop: minutes pass and gigabytes of intermediates accumulate before the error is revealed.} For the 1,046 partially built images (failed builds), the median intermediate size was 2.16 GB. Training partials were especially large (median 9.91 GB; max 114 GB), indicating that build overhead persists even on failure, burdening CI/CD systems and developer machines. In time, partial builds still consume non-trivial minutes: medians are 4.56 min (Training), 3.36 min (Sandbox), 1.72 min (Inference), and 1.08 min (Infrastructure), with long tails reaching 100+ min across roles (up to 125 min).

\begin{figure}[H]
\centering
\includegraphics[width=0.9\linewidth]{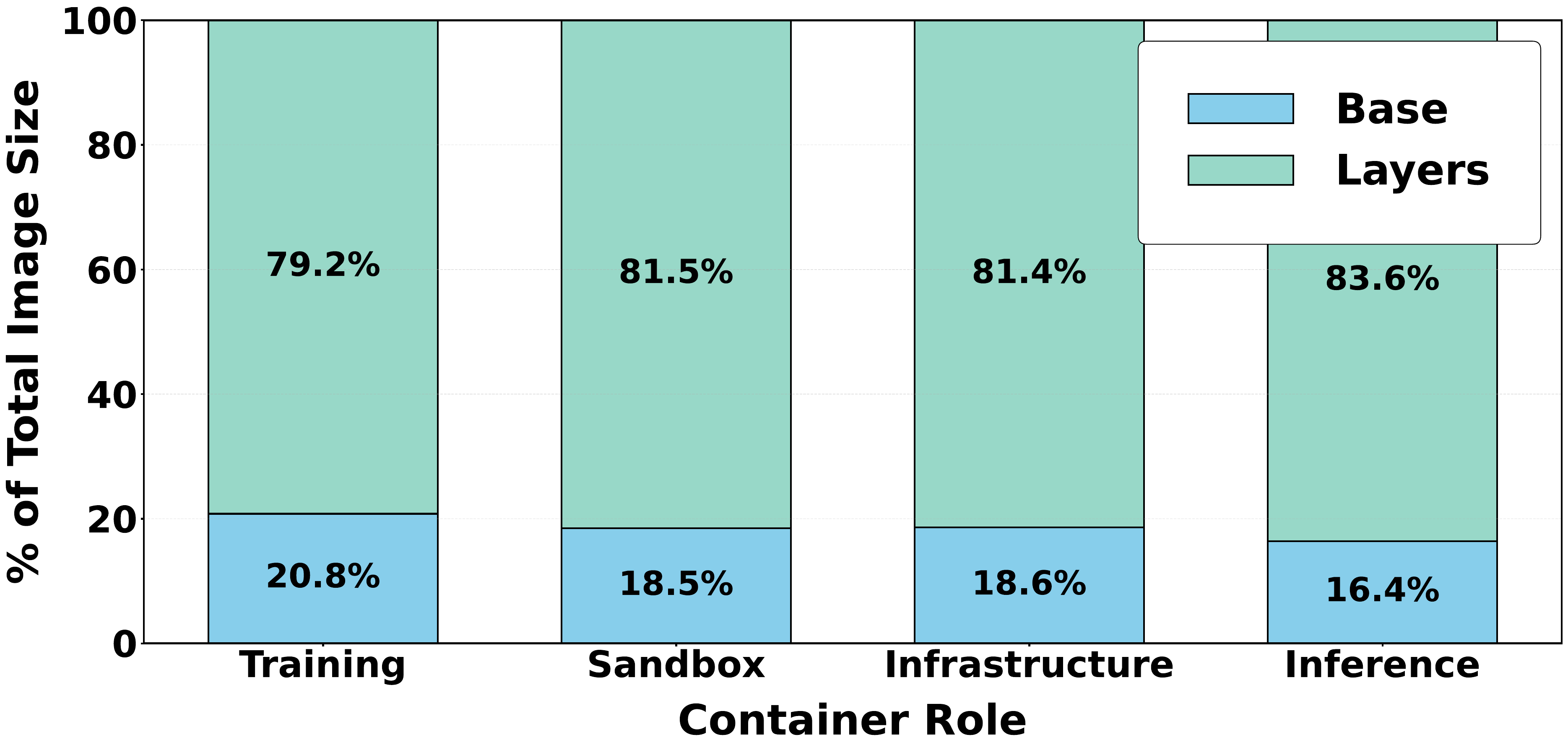}
\caption{Base \& Custom Layer Contributions to Image Size}
\label{fig:rq1_base_contribution}
\end{figure}

\textbf{Base images contribute little; custom layers hold most of the weight.} Across container types, the base image contributes only 16–21\% of the final size, leaving 79–84\% in project-added layers (\autoref{fig:rq1_base_contribution}; Training 20.8\%, Inference 16.4\%, Infrastructure 18.6\%, Sandbox 18.5\%). This pattern suggests most growth comes from project-added dependencies and artifacts, so optimization should focus on reorganizing or slimming custom layers.

\textbf{\texttt{RUN} layers dominate the custom footprint, \texttt{COPY}/\texttt{ADD} are substantial but secondary, and configuration is negligible.}
In Table~\ref{tab:rq1_layer_sizes_by_role}, the median \texttt{RUN} layer is the primary source of custom bytes across roles, 470~MB for training and 450~MB for sandbox, far above infrastructure (269~MB) and inference (228~MB), consistent with heavyweight installs during build. \texttt{COPY}/\texttt{ADD} layers are sizable but clearly smaller than \texttt{RUN}—173~MB (training), 170~MB (infrastructure), 127~MB (inference), and 119~MB (sandbox), reflecting the movement of code, models, and data into images. Configuration-only steps (\texttt{ENV}, \texttt{WORKDIR}, \texttt{CMD}, \texttt{ENTRYPOINT}) contribute effectively 0 MB because they modify metadata rather than filesystem content, yet each still creates a new image layer and therefore adds build-time.

\begin{table}[H]
\centering
\scriptsize
\setlength{\tabcolsep}{4pt}
\caption{Median size of custom layers per Dockerfile by instruction type.}
\label{tab:rq1_layer_sizes_by_role}
\begin{tabular}{lccc}
\toprule
\textbf{Container Role} & \textbf{RUN (MB)} & \textbf{COPY/ADD (MB)} & \textbf{Configuration (MB)} \\
\midrule
Training        & 470.00 & 172.91 & 0.00 \\
Sandbox         & 450.34 & 119.29 & 0.00 \\
Infrastructure  & 268.66 & 169.75 & 0.00 \\
Inference       & 228.01 & 127.17 & 0.00 \\
\bottomrule
\end{tabular}
\end{table}

\subsection{RQ2: Rebuild frequency, underlying causes, and the efficiency of cache reuse}

\begin{table}[h]
\centering
\setlength{\tabcolsep}{3pt}
\renewcommand{\arraystretch}{0.8}
\caption{Rebuild frequency and Trigger distribution}
\label{tab:rebuild_overview}
\scriptsize
\begin{tabular}{lcccccc}
\toprule
\textbf{Role} & 
\textbf{\shortstack{Rebuilt \\ Commits}} &
\textbf{\shortstack{Trigger:\\Copied files}} &
\textbf{\shortstack{Trigger:\\Dockerfile}} &
\textbf{\shortstack{Trigger:\\Both}} &
\textbf{\shortstack{Median \\ rebuild rate \\ perDockerfile}} \\
\midrule
Training   & 46.2\% \ (178{,}876/387{,}639) & 96.8\% & 1.0\% & 2.2\% & 19.5\% \\
Inference  & 43.4\% \ (290{,}295/669{,}639) & 91.7\% & 2.5\% & 5.9\% & 10.4\% \\
\midrule
\textbf{All} & 44.4\% \ (469{,}171/1{,}057{,}406) & 96.4\% & 1.1\% & 2.5\% & -- \\
\bottomrule
\end{tabular}
\end{table}

For the 939 Dockerfiles (632 inference + 307 training) from 221 repositories, we analyzed 1,057,406 commits since each Dockerfile was added; 44.4\% triggered a rebuild. Rebuild frequency and triggers distribution are summarized in (\autoref{tab:rebuild_overview}).

\textbf{Training containers are prone to rebuilds x2 as often as inference}. At the commit level, nearly half of training commits trigger a rebuild (46.2\%; 178,876/387,639), compared to 43.4\% for inference (290,295/669,639). Looking per Dockerfile, the gap widens: the median rebuild rate per Dockerfile is 19.5\% for training versus 10.4\% for inference. In other words, even though inference sees more total changes (commits), any given training change is more likely to force a rebuild.

\textbf{Most rebuilds begin outside the Dockerfile; context changes are the leading cause of image break.} Copied-file triggers account for 96.8\% of training rebuilds and 91.7\% of inference rebuilds, while Dockerfile-only triggers are rare (1.0\% training; 2.5\% inference). Simultaneous changes are also uncommon (2.2\% training; 5.9\% inference). Inference shows a larger Dockerfile-involved share overall (8.4\% vs. 3.2\% for training), but in both categories, the main driver remains regular updates to copied files (e.g., code, models, data, config files).

For all 452{,}281 (96.4\% of 469,171) rebuild commits triggered only by copied file edits, we computed the metric defined in (\autoref{sec:heuris}), as illustrated in \autoref{tab:cache_metrics} and \autoref{fig:rq1_base_contribution}.

\begin{figure}[H]
\centering
\includegraphics[width=0.75\linewidth]{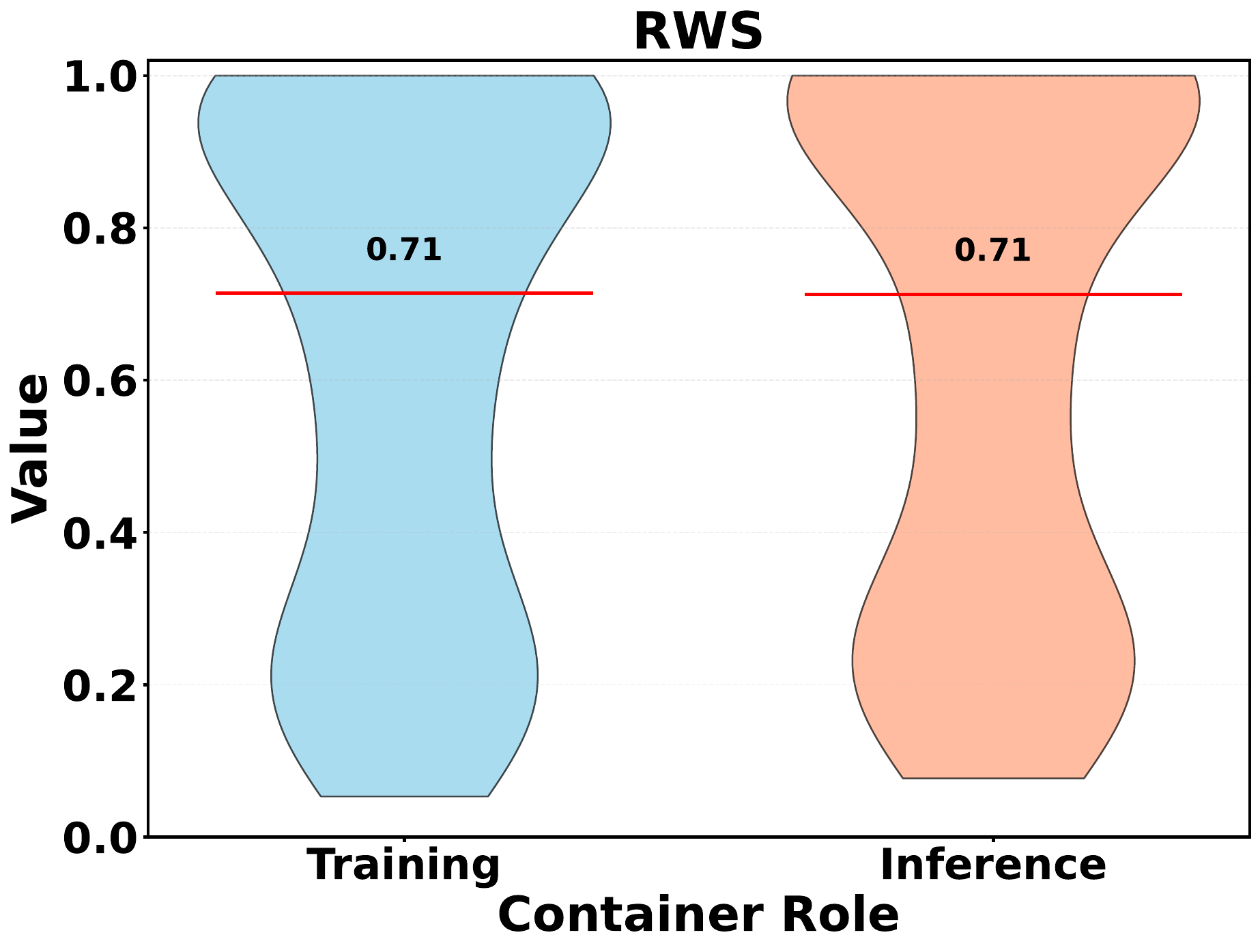}
\caption{Rebuild Work Share (RWS) Distribution}
\label{fig:rq1_base_contribution}
\end{figure}

\begin{table}[b]
\centering
\caption{Cache Reuse Share (CRS) and Cache-Break Depth (CBD) Metrics.}
\label{tab:cache_metrics}
\small
\begin{tabular}{lcccc}
\toprule
& \multicolumn{2}{c}{\textbf{CRS}} & \multicolumn{2}{c}{\textbf{CBD}} \\
\cmidrule(lr){2-3}\cmidrule(lr){4-5}
\textbf{Container Role} & \textbf{Median} & \textbf{[Q1, Q3]} & \textbf{Median} & \textbf{[Q1, Q3]} \\
\midrule
Training  & 0.286 & [0.048, 0.718] & 0.444 & [0.167, 0.714] \\
Inference & 0.287 & [0.000, 0.736] & 0.375 & [0.113, 0.700] \\
\bottomrule
\end{tabular}
\end{table}

\textbf{Only a third of cached work is typically reused, regardless of role.} In \autoref{tab:cache_metrics}, the median CRS is 0.286 for training and 0.287 for inference, with wide IQRs \([0.048,\,0.718]\) and \([0.000,\,0.736]\). Thus, a typical cache-breaking commit preserves about \(30\%\) of weighted work, but outcomes vary widely across projects. The lack of a material gap between roles suggests that, once the cache breaks, reuse depends more on Dockerfile structure (ordering and scope of heavyweight steps) than on whether the image trains or serves.

\textbf{Breaks arrive early and 70\% of the cached work is lost.} The median CBD places the first invalidated instruction around the mid-file (\autoref{tab:cache_metrics}), and the complementary distribution of RWS centers near \(0.7\) for both roles (\autoref{fig:rq1_base_contribution}), indicating that about two-thirds to three-quarters of weighted work is typically re-executed once the cache is invalidated. The shape of the distributions reinforces the point: early breaks dominate, late breaks are rare, and the resulting rebuild cost is both frequent and substantial.

\begin{figure}[htbp]
    \centering
    \includegraphics[width=0.8\linewidth]{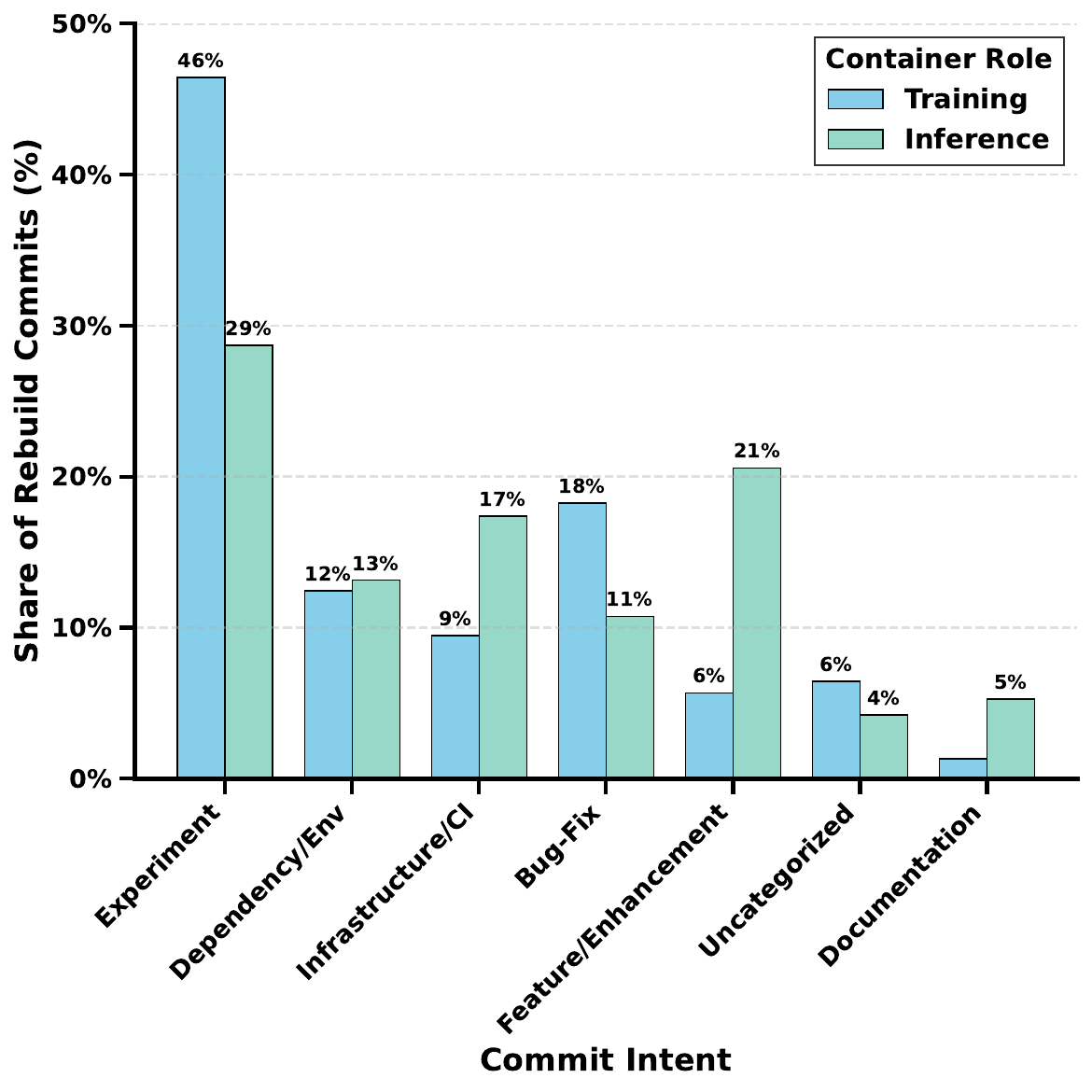}
    \caption{Rebuild commits Intent frequency}
    \label{fig:intent_frequency}
\end{figure}

A sample of 2,990 rebuild commits was categorized into six intent groups. 
\textbf{Training rebuilds are driven by experimental iteration, while inference rebuilds are driven by serving-code changes, infrastructure updates, and environment revisions.}
\autoref{fig:intent_frequency} shows a clear divergence in the commit intents that trigger container rebuilds. Training containers are dominated by \emph{Experiment} commits (46\%), reflecting edits to models, hyperparameters, and data assets, the core of iterative ML work. Inference containers, by contrast, exhibit a balanced composition in which \emph{Feature/Enhancement} changes to serving code (18\%), \emph{Infrastructure/CI} updates to pipelines (17\%), and \emph{Dependency/Environment} revisions to libraries, frameworks, CUDA, and base images (13\%) collectively outweigh \emph{Experiment} (29\%). A chi-square test confirms that these intent distributions differ significantly between roles ($\chi^2 = 282.96$, $p < 0.001$). Taken together, the patterns align with container purpose: training images serve fast-evolving research loops, whereas inference images evolve primarily through serving-side code evolution, pipeline/tooling changes, and environment updates.

\begin{figure}[htbp]
    \centering
    \includegraphics[width=\linewidth]{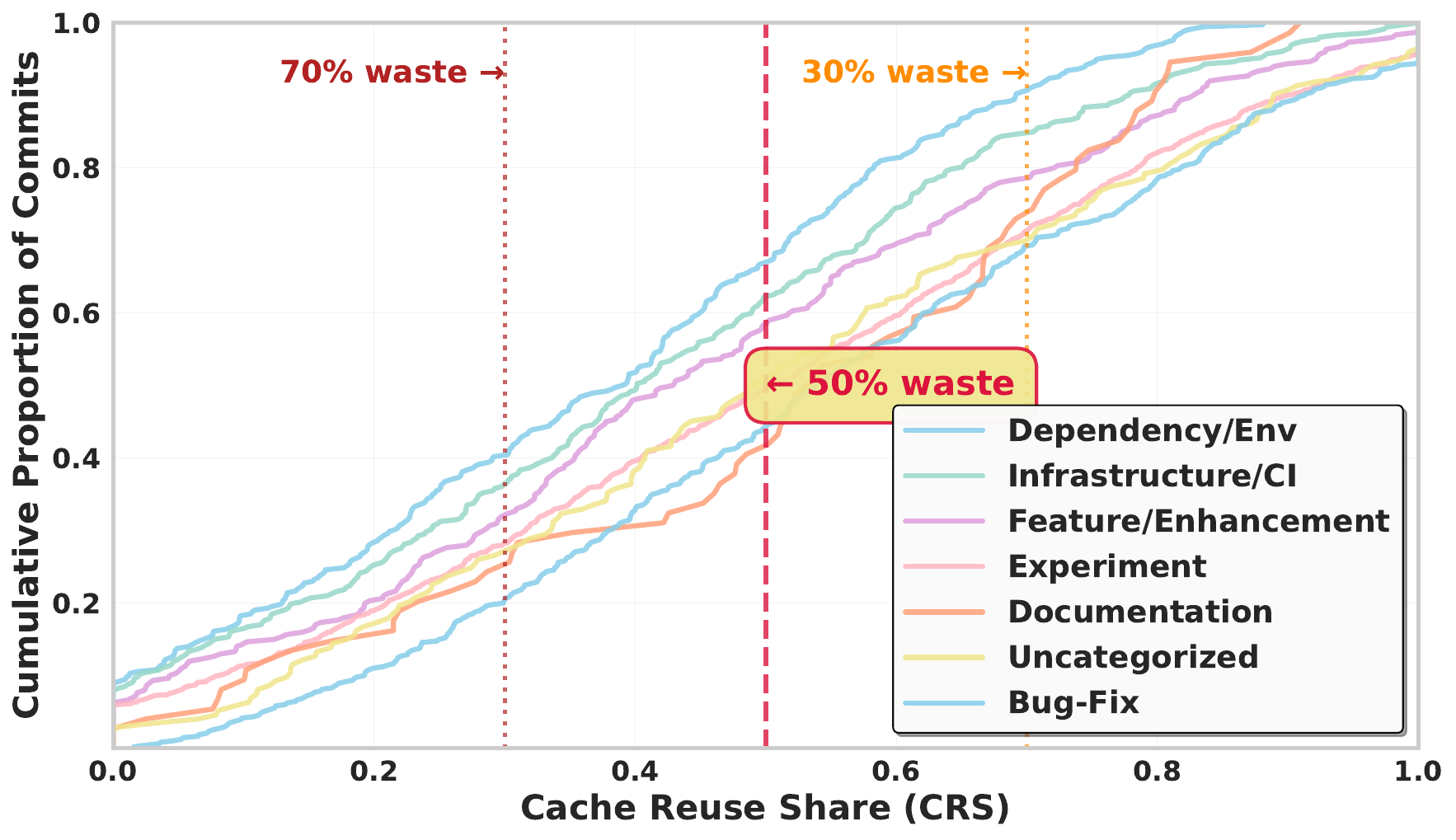}
    \caption{Cumulative distribution of CRS in rebuild commits}
    \label{fig:crs_cdf_intent}
\end{figure}

\autoref{fig:crs_cdf_intent} plots, for each intent category, the cumulative distribution function of the Cache Reuse Share (CRS) observed in rebuild-triggering commits. The x-axis is CRS ($0= \text{complete rebuild}$; $1=\text{full reuse}$). The y-axis is the cumulative fraction of commits with $\text{CRS} \le x$. Curves that rise early and steeply on the left indicate frequent low-reuse events (i.e., many commits near cold rebuilds). Curves that stay flat and shift right indicate high reuse (i.e., most commits preserve prior work). Vertical reference lines at $\text{CRS} \in \{0.3, 0.5, 0.7\}$ mark thresholds ($\approx 70\%/50\%/30\%$ of work redone).

\textbf{Cache reuse depends on commit intent; dependency and infrastructure modifications tend to trigger near-cold rebuilds.} \autoref{fig:crs_cdf_intent} shows the \emph{Dependency/Environment} and \emph{Infrastructure/CI} curves rising steeply in the low-CRS region: roughly half of these commits have $\mathrm{CRS} < 0.3$, signaling pervasive cache failure and near-cold rebuilds. Table~\ref{tab:intent_summary} confirms this with the lowest median scores ($\mathrm{CRS} 0.29$–$0.30$), implying that about $70\%$ of the weighted build work is typically re-executed after such changes. At the opposite end, \emph{Bug-Fix} (median $0.70$) and \emph{Documentation} ($0.65$) are right-shifted, indicating that most commits under these intents retain prior work. \emph{Experiment} ($0.57$) and \emph{Feature/Enhancement} ($ 0.44$) fall between these extremes, with broader CDF spread, consistent with mixed edit locality across projects.

\begin{table}[htbp]
\centering
\caption{Cache Reuse Share (CRS) and Aggregate Rebuild Cost by Commit Intent}
\label{tab:intent_summary}
\begin{adjustbox}{width=0.8\linewidth}
\begin{tabular}{lrr}
\toprule
\textbf{Intent} & 
\makecell{\textbf{Median CRS}\\\textbf{(role agnostic)}} & 
\makecell{\textbf{Global Cost}\\\textbf{Attribution \%}} \\
\midrule
Experiment         & 0.565 & 40.4 \\
Dependency / Env   & 0.288 & 15.3 \\
Bug-Fix            & 0.704 & 13.1 \\
Infrastructure / CI & 0.302 & 13.1 \\
Feature / Enhancement & 0.441 & 10.3 \\
Uncategorized      & 0.646 &  5.3 \\
Documentation      & 0.646 &  2.4 \\
\midrule
\textbf{Overall}   & \textbf{0.501} & \textbf{100.0} \\
\bottomrule
\end{tabular}
\end{adjustbox}

\begin{tablenotes}\footnotesize
\item Global Cost Attribution $= \sum_{i \in \text{intent}}\mathrm{RWS}_i / \sum_{\text{all}}\mathrm{RWS}$ (proportion of total wasted build work).
\end{tablenotes}
\end{table}

\begin{table*}[t]
\centering
\scriptsize
\setlength{\tabcolsep}{3.5pt}
\renewcommand{\arraystretch}{1.12}
\caption{ML-Specific Dockerfile Refactorings}
\label{tab:rq3_patterns_refactorings_compact}
\begin{tabularx}{0.98\linewidth}{p{30mm} p{12mm} X p{20mm} p{15mm} r}
\toprule
\textbf{Refactoring (Name)} & \textbf{Container Role} & \textbf{Definition (action)} & \textbf{Target} & \textbf{Example (link/commit)} & \textbf{Count} \\
\midrule

\textbf{Externalize Dataset Artifact} &
Training &
Remove dataset \texttt{COPY}/downloads from the Dockerfile and load datasets at runtime via a mounted volume or startup fetch from a versioned store. &
Size $\downarrow$, (Re)Build Time $\downarrow$ &
\href{https://github.com/jruokola/llm-granite-ft/commit/7f5055d993f731359ad6bb606219975876f7e36e}{7f5055d}  & 7 \\
\midrule

\textbf{Externalize Model Weights} &
Inference &
Delete model-weight \texttt{COPY}/fetch and resolve weights at runtime via bind-mount or registry download. &
Size $\downarrow$, (Re)Build Time $\downarrow$ &
\href{https://github.com/inarighas/voice-transcript/commit/7d47b14d3bfd69a28489e30c0b2793a7b7da3d1a}{7d47b14}  & 5 \\
\midrule

\textbf{Reallocate Model Downloads} &
Training &
Move deterministic downloads of pre-trained/base models from late stages to an early RUN layer so the artifact is cached across code/config edits. &
(Re)Build Time $\downarrow$ &
\href{https://github.com/UMassCDS/inatator/commit/fc2c244e8c04b5decd700356a935c971000dfdc6}{fc2c244} & 5 \\
\midrule

\textbf{Split Container Image by Accelerator (CPU/GPU)} &
Both &
When both CPU and GPU targets are required, split the unified image into two Dockerfiles (Dockerfile.cpu and Dockerfile.gpu), each using its own base, and remove the non-target stacks. &
Size $\downarrow$, (Re)Build Time $\downarrow$ &
\href{https://github.com/MetrodataTeam/sherpa-grpc/commit/b65919e643b7c5e1bc0c8c8e3c535e0cafdc9fc0}{b65919e}  & 8 \\
\midrule

\textbf{Select Hardware-Targeted ML Framework} &
Both &
Install only the ML framework variant that matches the target hardware, align drivers/base to its compatibility, and exclude alternate variants. &
Size $\downarrow$, (Re)Build Time $\downarrow$ &
\href{https://github.com/Team-Troef-Energy/SeriousGame/commit/5621177024e27e882b45eaf4a3dfbdab49320a51}{5621177} & 4 \\
\midrule

\textbf{Reallocate Model Configuration} &
Training &
Reorder the Dockerfile to copy frequently edited configs/scripts (used for finetuning) after heavyweight installs so frequent edits do not invalidate earlier layers. &
(Re)Build Time $\downarrow$ &
\href{https://github.com/ahmed-n-abdeltwab/spyware-detector-training/commit/f3db2c01fec808a182551e525e4e1625e5aaa9a3}{f3db2c0}  & 4 \\
\midrule

\textbf{Shallow-Clone VCS (Version Control System) Dependencies} &
Both &
When using git clone for dependency download, replace full-history \texttt{git clone} with \texttt{--depth} (and optional sparse checkout) pinned to a commit SHA. The same source snapshot is built with less network and unpack work. &
Build Time $\downarrow$, Size $\downarrow$ &
\href{https://github.com/ai4prod/aimet_cuda_12/commit/7bb8b0735720e47b5184c88b6bbfe59cc11ea5ba}{7bb8b07} & 7 \\
\bottomrule
\end{tabularx}
\end{table*}

\textbf{Experiments are the primary source of global cache waste, mainly due to their prevalence.}
\autoref{tab:intent_summary}’s global cost Attribution combines frequency with cache reuse to quantify each intent’s share of total wasted build work. \emph{Experiment} commits contribute 40.4\%,the single largest portion, because they occur often (especially in training), even though their median CRS is only moderate. \emph{Dependency/Environment} and \emph{Infrastructure/CI} add 15.3\% and 13.1\%, respectively; their overall impact stems from fewer but costly events. Notably, \emph{Bug-Fix} changes also account for 13.1\%, underscoring that high-reuse intents can still matter globally when they are common. Taken together, experiments, dependency updates, and infrastructure changes comprise over two-thirds of wasted rebuild work.

\subsection{RQ3: ML-specific Dockerfile Refactorings.}

Across 29 projects and 377 Dockerfile-changing commits, we found 40 behavior-preserving refactorings that cluster into 7 patterns (Table \ref{tab:rq3_patterns_refactorings_compact}). Four patterns apply to both training and inference; all explicitly target image size and/or (re)build time. Below we define each pattern concisely and note observed counts along with an example from real commits.

\textbf{Externalize Dataset Artifact:} Many training images balloon because datasets are copied or downloaded during build; any data update then forces a rebuild and ships gigabytes to registries. This refactoring removes dataset COPY/download steps from the Dockerfile and binds data at runtime (mounted volume or startup fetch from a versioned store with checksums). It preserves behavior for the same dataset version but shrinks footprint (RQ1: fewer large COPY layers) and eliminates rebuilds on data iteration (RQ2: cache bypassed since the artifact is no longer part of the image). We observed this in 7 commits. \emph{Example.} In \href{https://github.com/jruokola/llm-granite-ft/commit/7f5055d993f731359ad6bb606219975876f7e36e}{commit~7f5055d}
(jruokola/-llm-granite-ft), the Dockerfile drops dataset \texttt{COPY} steps and introduces a startup script that fetches the dataset at container launch.

 \textbf{Externalize Model Weights:} Inference images often embed model files (hundreds of MB–GB), coupling model rollout to image rebuilds. This refactoring deletes model‐weight COPY/fetch from the Dockerfile and loads weights at startup (bind-mount or registry download. With this practice, image size decreases and model iterations no longer trigger rebuilds; We observed 5 instances of this refactoring. \emph{Example.} In \href{https://github.com/inarighas/voice-transcript/commit/7d47b14d3bfd69a28489e30c0b2793a7b7da3d1a}{commit~7d47b14}
 (inarighas/-voice-transcript), the Dockerfile stops copying model files and instructs to mount pretrained models instead of copying them.

\textbf{Reallocate Model Downloads:} When pre-trained/base model artifacts are stable over many commits (e.g., a fixed base checkpoint for fine-tuning), placing the download late in the Dockerfile causes re-download after small code/config edits. This refactoring moves the pinned download (URL/SHA) to an early RUN layer, keeping the version constant so the layer is cacheable across iterations. We observed 5 instances of this pattern. \emph{Example.} In \href{https://github.com/UMassCDS/inatator/commit/fc2c244e8c04b5decd700356a935c971000dfdc6}{commit~fc2c244}
(UMassCDS/-inatator), the author explicitly notes: “\emph{Moved the pre-trained model download to the start of the build to prevent repeated downloads during rebuilds},” and the diff shows the model-download \texttt{RUN} command moved from a late stage to the top of the Dockerfile.

\textbf{Split Container Image by Accelerator (CPU/GPU):}
Unified images carrying both CPU and CUDA stacks are large and fragile to invalidation. When supporting both targets, this refactoring materializes two images, Dockerfile.cpu and Dockerfile.gpu, each with a matching base and non-target stacks removed (drop CUDA/cuDNN in the CPU image; drop CPU-only extras from the GPU image). The application behavior on each target is preserved, while custom bytes fall (fewer heavyweight RUN installs; RQ1) and rebuild scope shrinks (fewer early layers to invalidate; RQ2). This was the most common pattern (8 commits). \emph{Example.} In \href{https://github.com/MetrodataTeam/sherpa-grpc/commit/b65919e643b7c5e1bc0c8c8e3c535e0cafdc9fc0}{commit~b65919e}
 (MetrodataTeam/-sherpa-grpc), the author separates CPU and GPU Dockerfiles (commit title: “\emph{separate cpu/gpu dockerfile}”), introducing a minimal CPU base for the CPU image and a CUDA runtime base for the GPU image, eliminating non-target stacks in each.

\textbf{Select Hardware-Targeted ML Framework:} Framework installations that omit hardware qualifiers (e.g., installing PyTorch without the -cpu suffix) often pull full distributions containing unused GPU binaries, inflating image size and increasing solver and unpack work. This refactoring selects the hardware-specific distribution of the ML framework, aligns the base image and drivers with its compatibility matrix, and excludes non-target variants. The result is a leaner dependency stack (smaller RUN layers; RQ1) and reduced rebuild work when dependencies change (fewer massive reinstall steps; RQ2). We observed 4 instances of this pattern.
\emph{Example.} In \href{https://github.com/Team-Troef-Energy/SeriousGame/commit/5621177024e27e882b45eaf4a3dfbdab49320a51}{commit~5621177} (Team-Troef-Energy/SeriousGame), the author explicitly switches to a CPU-only framework build, stating: "\emph{to reduce the image size from 1 GB, use torch+cpu to avoid pulling in…}" This change concretely demonstrates hardware-targeted package selection to eliminate unnecessary CUDA artifacts.

\textbf{Reallocate Model Configuration: } Placing frequently changing files (fine-tuning config files) above dependency installations invalidates the bulk of the build with every experiement. This refactoring reorders the Dockerfile so these files are copied after framework/toolchain installation. Resulting in faster rebuilds with no footprint change. We found 4 cases. \emph{Example.} In \href{https://github.com/ahmed-n-abdeltwab/spyware-detector-training/commit/f3db2c01fec808a182551e525e4e1625e5aaa9a3}{commit~f3db2c0}, the Dockerfile replaces a coarse \texttt{COPY . .} with explicit \texttt{COPY}s (\texttt{setup.py}, \texttt{requirements.txt}, \texttt{src/}, and \texttt{config/}) placed after framework/toolchain installation. This isolates volatile files such as \texttt{config/../model\_trainer.yml} to late layers, so tweaks to training configs no longer invalidate earlier dependency layers.

\textbf{Shallow-Clone VCS Dependencies:}
Full-history git clone of large ML dependencies dominates network and unpack time and inflates layers with unnecessary history. This refactoring replaces full clones with \texttt{--depth} (and optional sparse checkout) pinned to a commit SHA, leaving downstream build steps unchanged. It reduces cold build time and shrinks custom layers, without altering the source snapshot the image compiles against. We observed 7 cases of this refactoring. \emph{Example.} In \href{https://github.com/ai4prod/aimet_cuda_12/commit/7bb8b0735720e47b5184c88b6bbfe59cc11ea5ba}{commit~7bb8b07}
 (ai4prod/-aimet\_cuda\_12), The Dockerfile originally cloned Qualcomm’s AIMET repository using a full-history git clone, which downloaded the entire commit log and submodules. The refactoring modifies this command to use  \texttt{git clone --depth 1}, retrieving only the latest snapshot pinned to a specific commit.
\section{Discussion}
\label{sec:discussion}
Our findings imply that ML containerization must evolve from ad hoc reuse of general-purpose Docker practices toward ML-specific discipline. Teams should externalize large artifacts such as datasets and model weights instead of embedding them, reorder Dockerfiles to install stable ML dependencies before copying volatile experiment code, and separate CPU/GPU targets into distinct builds to reduce churn and image footprint. Since base images contribute little to total size, optimization efforts should target custom layers and build ordering rather than distro minimization. Our heuristic metrics (CRS/RWS) can guide practitioners in estimating the rebuild cost of edits without executing builds, supporting proactive pipeline tuning. Together, these insights call for ML-aware CI/CD tooling, layer-sensitive caches, data-versioned artifact mounts, and static analyzers that flag cache-fragile Dockerfile layouts, to make experimentation faster, more sustainable, and reproducible.

\section{Threats to Validity}
We now discuss the threats to the validity of our study following the guidelines for case study research~\cite{yin2009case}.

\emph{\textbf{Threats to Construct Validity}} Mislabeling container roles (e.g., training vs. inference vs. infrastructure) could distort role-specific build and size comparisons. We mitigated this through a structured multi-rater process: an initial calibration batch, double-coding of 25\% of files, and consensus reconciliation, achieving substantial inter-rater agreement (Cohen’s K = 0.72–0.83, Fleiss’ K= 0.78). Ambiguous multi-purpose containers were conservatively assigned the dominant role based on entrypoints and invoked scripts. Build duration can conflate compute, I/O, and network latency. To reduce noise, we standardized the hardware, ran each build three times, and averaged the results. Image footprint was measured as the total uncompressed size of all layers. The cache-reuse (CRS/RWS) heuristic approximates Docker’s rebuild cost without executing historical builds. Threats include imperfect weighting of instruction types and unmodeled differences in package manager behavior, which we addressed by basing the weight schedule on Docker’s documented cache semantics and the observed empirical layer characteristics in RQ1. LLM-based labeling can introduce noise; to mitigate this, inputs were limited to commit messages and relevant file diffs, and results were audited (79\% primary, 89\% overlap).

\emph{\textbf{Threats to Internal Validity}}
Build caching state can influence timing. We clearly distinguish cold builds (RQ1) from heuristic hot-cache scenarios (RQ2). For RQ1, caches were flushed between runs. Given high real-world Dockerfile failure rates (\cite{sakamoto2025automatic, Ksontini2025MSR}), we mitigate survivorship bias by retaining partial build signals (time-to-failure, produced-layer size) and incorporating them into summaries. For repository history analysis, our linearized (first-parent) history avoids double-counting merges but may miss cache invalidations in long-lived side branches. Since our focus is on the mainline development path where releases and CI builds occur, the effect on conclusions is minimal. applying each project’s .dockerignore reduces false positives when detecting copied-files triggers.

\emph{\textbf{Threats to External Validity}}
Our corpus is drawn from recent, publicly available, Python-based ML repositories covering end-to-end pipelines. This focuses on production projects but limits generalization to Python ecosystems and public codebases. Practices in enterprise monorepos, closed-source environments, or non-Python ML stacks (e.g., Java) may differ. We study standard Dockerfiles built under Docker/BuildKit semantics. Other systems (Bazel, e.g., podman ) may exhibit different caching behavior. Nevertheless, the principles we identify, such as isolating dependencies and minimizing early cache breaks, apply across container platforms.

\section{Conclusion}
\label{sec:conclusion}

This study provides a role-aware, data-driven view of ML container practice at scale. From 1,993 Dockerfiles across 392 repositories, we show that footprint and cold-build latency concentrate by role: training images are large (median 17 GB) and slow (15 min), while inference and infrastructure are smaller and faster (medians 1.5–1.7 GB and 1–1.3 min). Base layers account for a minority of bytes (16–21\%), with most growth in custom layers (heavy RUN installs and sizable COPYs). Longitudinally, about 44\% of commits trigger a rebuild, driven by context edits (>90\%), and once the cache breaks, typical reuse is low (median Cache Reuse Share 0.29), implying that ~70\% of work is redone. An examination of commit histories reveals that the nature of changes is important. Training rebuilds are predominantly characterized by Experiment edits, whereas inference rebuilds more frequently result from modifications related to Dependency/Environment, Infrastructure/CI, and Feature/Enhancement, which impact foundational, resource-intensive layers and lead to near-cold rebuilds. We also observe that developers actively refactor their ML Dockerfiles. From real commits, we identified seven recurring, behavior-preserving patterns.

For researchers, our study contributes a publicly grounded role taxonomy and dataset, operational metrics for rebuild analysis, and an empirically derived catalog of refactorings aligned with observed failure modes. Together, these provide baselines and measurement instruments for evaluating ML-aware build techniques and intent-sensitive caching.

\bibliographystyle{ACM-Reference-Format}
\bibliography{references}

\end{document}